\documentclass{article}
\usepackage{amsmath,amssymb,enumerate,bbm,hyperref}

\newcommand{\nocomma}{}
\newcommand{\tmem}[1]{{\em #1\/}}
\newcommand{\tmop}[1]{\ensuremath{\operatorname{#1}}}
\newcommand{\tmstrong}[1]{\textbf{#1}}
\newcommand{\tmtextbf}[1]{{\bfseries{#1}}}
\newenvironment{enumerateroman}{\begin{enumerate}[i.] }{\end{enumerate}}
\newenvironment{itemizedot}{\begin{itemize} }{\end{itemize}}
\newtheorem{lemma}{Lemma}
\newtheorem{proposition}{Proposition}
\newtheorem{theorem}{Theorem}

\begin{document}

\title{Polynomial functions of degree 20 which are APN infinitely
often.}\author{Florian Caullery \\ Institut of Mathematiques of Luminy \\ C.N.R.S. \\ France \\ florian.caullery@etu.univ-amu.fr }\maketitle

\begin{abstract}
  We give all the polynomials functions of degree 20 which are APN over an
  infinity of field extensions and show they are all CCZ-equivalent to the
  function $x^5$, which is a new step in proving the conjecture of Aubry,
  McGuire and Rodier.
\end{abstract}

{\tmstrong{Keywords: }}vector Boolean functions, almost perfect nonlinear
functions, algebraic surface, CCZ-equivalence.

\section{Introduction}

Modern private key crypto-systems, such as AES, are block cipher. The security
of such systems relies on what is called the S-box. This is simply a Boolean
function $f : \mathbbm{F}_{2^n} \rightarrow \mathbbm{F}_{2^n}$ where $n$ is
the size of the blocks. It is the only non linear operation in the algorithm.

One of the best known attack on these systems is differential cryptanalysis.
Nyberg proved in [13] that the S-boxes with the best resistance to such
attacks are the one who are said to be Almost Perfectly Non-linear (APN).

Let $q = 2^n$. A function $f : \mathbbm{F}_q \rightarrow \mathbbm{F}_q$ is
said APN on $\mathbbm{F}_q$ if the number of solutions in $\mathbbm{F}_q$ of
the equation
\[ f \left( x + a \right) + f \left( x \right) = b \]
is at most 2 for all $a, b \in \mathbbm{F}_q$, $a \neq 0$. The fact that
$\mathbbm{F}_q$ has characteristic 2 implies that the number of solutions is
even for any function $f$ on $\mathbbm{F}_q$.

The study of APN functions has focused on power functions and it was recently
generalized to other functions, particularly polynomials (Carlet, Pott and al.
[5, 7, 8]) or polynomials on small fields (Dillon [6]). On the other hand,
several authors (Berger, Canteaut, Charpin, Laigle-Chapuy [2], Byrne, McGuire
[4] or Jedlicka [10]) showed that APN functions did not exist in certain
cases. Some also studied the notion of being APN on other fields than
$\mathbbm{F}_{2^n}$ (Leducq [12]).

Toward a full classification of all APN functions, an approach is to show that
certain polynomials are not APN for an infinity of extension of
$\mathbbm{F}_2$.

Hernando and McGuire showed a result on classification of APN functions which
was conjectured for 40 years : the only exponents such that the monomial $x^d$
is APN over an infinity of extension of $\mathbbm{F}_2$ are of the form $2^i +
1$ or $4^i - 2^i + 1$. Those exponents are called {\tmem{exceptional
exponents}}.

It lead Aubry, McGuire and Rodier to formulate the following conjecture:

{\tmstrong{Conjecture:}} (Aubry, McGuire and Rodier) a polynomial can be APN
for an infinity of ground fields $\mathbbm{F}_q$ if and only if it is
CCZ-equivalent (as defined by Carlet, Charpin and Zinoviev in [5]) to a
monomial $x^d$ where $d$ is an exceptional exponent.

A way to prove this conjecture is to remark that being APN is equivalent to
the fact that the rational points of a certain algebraic surface $X$ in a
3-dimensional space linked to the polynomial $f$ defining the Boolean function
are all in a surface $V$ made of 3 planes and independent of $f$. We define
the surface $X$ in the 3-dimensional affine space $\mathbbm{A}^3$ by
\[ \phi \left( x, y, z \right) = \frac{f \left( x \right) + f \left( y \right)
   + f \left( z \right) + f \left( x + y + z \right)}{\left( x + y \right)
   \left( x + z \right) \left( y + z \right)} \]
which is a polynomial in $\mathbbm{F}_q \left[ x, y, z \right]$. When the
surface is irreducible or has an irreducible component defined over the field
of definition of $f$, a Weil's type bound may be used to approximate the
number of rational points of this surface. When it is too large it means the
surface is too big to be contained in the surface $V$ and the function $f$
cannot be APN.

This way enabled Rodier to prove in [14] that when the degree of $f$ is equal
to $4 e$ with $e \equiv 3 \left( \tmop{mod} 4 \right)$ and $\phi$ is not
divisible by a certain form of polynomial then $f$ is not APN for an infinity
of extension of $\mathbbm{F}_q$. He also found all the APN function of degree
12 and proved they are all CCZ-equivalent to $x^3$.

To continue in this way, let's get interested in the APN functions of degree
20 which were the next ones on the list. The main difference in this case is
that $e \equiv 1 \left( \tmop{mod} 4 \right)$. We got inspired by the proof of
Rodier in [14] but we had an other approach using divisors of the surface
$\bar{X}$. This was due to the fact that some of the components of $\bar{X}$
are no longer irreducible in our case.

Then we were able to obtain all the APN functions of degree 20 by calculation.
The conditions of divisibility by the polynomials we obtained made the first
part of our work, we had to work on the quotient after to obtain the final
forms of the functions.

The second part was to prove that all were CCZ-equivalent to $x^5$.

{\tmem{This work has been done with Fran\c{c}ois Rodier as adviser.}}

\section{The state of the art}

The best known APN functions are the Gold functions $x^{2^i + 1}$ and the
Kasami-Welch functions by $x^{4^i - 2^i + 1}$. These 2 functions are defined
over $\mathbbm{F}_2$ and they are APN on any field $\mathbbm{F}_{2^m}$ if
$\gcd \left( m, i \right) = 1$. Aubry, McGuire and Rodier obtained the
following results in [1].

\begin{theorem}
  {\tmstrong{(Aubry, McGuire and Rodier, [1])}} If the degree of the
  polynomial function f is odd and not an exceptional number then f is not APN
  over $\mathbbm{F}_{q^n}$ for all n sufficiently large.
\end{theorem}

\begin{theorem}
  {\tmstrong{(Aubry, McGuire and Rodier [1])}} If the degree of the polynomial
  function f is 2e with e odd and if f contains a term of odd degree, then f
  is not APN over $\mathbbm{F}_{q^n}$ for all n sufficiently large.
\end{theorem}

There are some results in the case of Gold degree $d = 2^i + 1$:

\begin{theorem}
  {\tmstrong{(Aubry, McGuire and Rodier [1])}} Suppose $f \left( x \right) =
  x^d + g \left( x \right)$ where $\deg \left( g \right) \leqslant 2^{i - 1} +
  1$. Let $g \left( x \right) = \sum_{j = 0}^{2^{i - 1} + 1} a_j x^j$. Suppose
  moreover that there exists a nonzero coefficient $a_j$ of g such that
  $\phi_j \left( x, y, z \right)$ is absolutely irreducible (where $\phi_j
  \left( x, y, z \right)$ denote the polynomial $\phi \left( x, y, z \right)$
  associated to $x^j$). Then f is not APN over $\mathbbm{F}_{q^n}$ for all n
  sufficiently large.
\end{theorem}

And for Kasami degree as well:

\begin{theorem}
  {\tmstrong{(F\'erard, Oyono and Rodier [9])}} Suppose $f \left( x \right) =
  x^d + g \left( x \right)$ where $d$ is a Kasami exponent and $\deg \left( g
  \right) \leqslant 2^{2 k - 1} - 2^{k - 1} + 1$. Let $g \left( x \right) =
  \sum_{j = 0}^{2^{2 k - 1} - 2^{k - 1} + 1} a_j x^j .$ Suppose moreover that
  there exist a nonzero coefficient $a_j$ of g such that $\phi_j \left( x, y,
  z \right)$ is absolutely irreducible. Then $\phi \left( x, y, z \right)$ is
  absolutely irreducible.
\end{theorem}

Rodier proved the following results in [14]. We recall that for any function
$f : \mathbbm{F}_q \rightarrow \mathbbm{F}_q$ we associate to $f$ the
polynomial $\phi \left( x, y, z \right)$ defined by:
\[ \phi \left( x, y, z \right) = \frac{f \left( x \right) + f \left( y \right)
   + f \left( z \right) + f \left( x + y + z \right)}{\left( x + y \right)
   \left( x + z \right) \left( y + z \right)} . \]
\begin{theorem}
  {\tmstrong{(Rodier [14])}} If the degree of a polynomial function f is even
  such that $\deg \left( f \right) = 4 e$ with $e \equiv 3 \left( \tmop{mod} 4
  \right)$, and if the polynomials of the form
  \[ \left( x + y \right) \left( x + z \right) \left( y + z \right) + P, \]
  with
  \[ P \left( x, y, z \right) = c_1 \left( x^2 + y^2 + z^2 \right) + c_4
     \left( xy + xz + zy \right) + b_1 \left( x + y + z \right) + d, \]
  for $c_1, c_4, b_1, d \in \mathbbm{F}_{q^3}$, do not divide $\phi$ then $f$
  is APN over $\mathbbm{F}_{q^n}$ for $n$ large.
\end{theorem}

There are more precise results for polynomials of degree 12.

\begin{theorem}
  {\tmstrong{(Rodier [14])}} If the degree of the polynomial f defined over
  $\mathbbm{F}_q$ is 12, then either f is not APN over $\mathbbm{F}_{q^n}$ for
  large n or f is CCZ equivalent to the Gold function $x^3$.
\end{theorem}

\section{New Results}

We have been interested in the functions defined by a polynomial of degree 20.

The main difference with the case already studied is that, when $e = 5$,
$\phi_e \left( x, y, z \right)$ (where $\phi_e \left( x, y, z \right)$ denote
the polynomial $\phi \left( x, y, z \right)$ associated to $x^e$) is not
irreducible. So we had to detail more cases in the proof and use divisors on
the surface $X$. And then obtained the following results :

\begin{theorem}
  If the degree of a polynomial function defined over $\mathbbm{F}_q$ is 20
  and if the polynomials of the form
  \[ (x + y) (x + z) (y + z) + P_1 \]
  with $P_1 \in \mathbbm{F}_{q^3} \left[ x, y, z \left] \right. \right.$ and
  $P_1 (x, y, z) = c_1 (x^2 + y^2 + z^2) + c_4 (xy + xz + yz) + b_1 (x + y +
  z) + d$

  or
  \[ \phi_5 + P_2 \]
  with $P_2 = a (x + y + z) + b$

  do not divide $\phi$ then $f$ is APN over $\mathbbm{F}_{q^n}$ for $n$ large.
\end{theorem}

\begin{theorem}
  If the degree of the polynomial f defined over $\mathbbm{F}_q$ is 20, then
  either f is not APN over $\mathbbm{F}_{q^n}$ for large n or f is CCZ
  equivalent to the Gold function $x^5$.
\end{theorem}

\section{Preliminaries}

The following results are needed to prove the theorem 7 All the proofs are in
[14].

\begin{proposition}
  \tmtextbf{[14]}The class of APN functions is invariant by adding a q-affine
  polynomial.
\end{proposition}

\begin{proposition}
  \tmtextbf{[14]}The kernel of the map
  \[ f \rightarrow \frac{f \left( x \right) + f \left( y \right) + f \left( z
     \right) + f \left( x + y + z \right)}{\left( x + y \right) \left( x + z
     \right) \left( y + z \right)} \]
  is made of q-affine polynomials.
\end{proposition}

We define the surface $X$ in the 3-dimensional affine space $\mathbbm{A}^3$ by
\[ \phi \left( x, y, z \right) = \frac{f \left( x \right) + f \left( y \right)
   + f \left( z \right) + f \left( x + y + z \right)}{\left( x + y \right)
   \left( x + z \right) \left( y + z \right)} \]
and we call $\bar{X}$ its projective closure.

\begin{proposition}
  \tmtextbf{[14]}If the surface X has an irreducible component defined over
  the field of definition of $f$ which is not one of the planes $\left( x + y
  \right) \left( x + z \right) \left( y + z \right) = 0$, the function $f$
  cannot be APN for infinitely many extensions of $\mathbbm{F}_q$.
\end{proposition}

\begin{lemma}
  \tmtextbf{[11]}Let $H$ be a projective hyper-surface. If $\bar{X} \cap H$
  has a reduced absolutely irreducible component defined over $\mathbbm{F}_q$
  then $\bar{X}$ has an absolutely irreducible component defined over
  $\mathbbm{F}_q$.
\end{lemma}

\begin{lemma}
  \tmtextbf{[1]}Suppose $d$ is even and write $d = 2^j e$ where $e$ is odd. In
  $\bar{X} \cap H$ we have
  \[ \phi_d = \phi_e \left( x, y, z \right)^{2^j} \left( \left( x + y \right)
     \left( x + z \right) \left( y + z \right) \right)^{2^j \text{-} 1} \]
\end{lemma}

\begin{lemma}
  \tmtextbf{[14]}The function $x + y$ (and therefore $A$) does not divide
  $\phi_i \left( x, y, z \right)$ for $i$ an odd integer.
\end{lemma}

\begin{lemma}
  {\tmem{$\phi_5$}} is not irreducible and we have
  \[ \phi_5 = \left( x + \alpha y + \alpha^2 z \right) \left( x + \alpha^2 y +
     \alpha z \right) \]
  with $\alpha \in \mathbbm{F}_4 \text{-} \mathbbm{F}_2$.
\end{lemma}

Calculus is sufficient to prove this.

\section{Proof of theorem 7}

Let $f : \mathbbm{F}_q \rightarrow \mathbbm{F}_q$ be a function which is APN
over infinitely many extensions of $\mathbbm{F}_q$. As a consequence of
proposition 11 no absolutely irreducible component of $X$ is defined over
$\mathbbm{F}_q$, except perhaps $x + y = 0$, $x + z = 0$ or $y + z = 0$.

If some component of $X$ is equal to one of these planes then by symmetry in
$x \nocomma$, $y$, and $z$, all of them are component of $X$, which implies
that $A = \left( x + y \right) \left( x + z \right) \left( y + z \right)$
divides $\phi$. Let us suppose from now on that this is not the case.

Let $H_{\infty}$ is the plane at infinity of $\mathbbm{A}^3$ and $X_{\infty} =
\bar{X} \cap H_{\infty}$. The equation of $X_{\infty}$ is $\phi_{20} = 0$
which gives, using lemma 13 and 14
\[ A^3 \left( x + \alpha y + \alpha^2 z \right)^4 \left( x + \alpha^2 y +
   \alpha z \right)^4 = 0 \]
As the curve $X_{\infty}$ does not contain any irreducible component defined
over $\mathbbm{F}_q$, $\alpha \notin \mathbbm{F}_q$ and then $q = 2^n$ with
$n$ odd.

Let $X_0$ be a reduced absolutely irreducible component of $\bar{X}$ which
contains the line $x + y = 0$ in $H_{\infty}$. The cases where $X_0$ contains
2 or 3 copies of the line $x + y = 0$ in $H_{\infty}$ and where $X_0$ contains
one copy of the line $x + y = 0$ and is of degree $1$ are treated in [14] and
do not differ in our case. So from now on we assume that $X_0$ contains only
one copy of the line $x + y = 0$ and is at least of degree 2.

Let $d_1$ be the plane of equation $\left( x + \alpha y + \alpha^2 z \right) =
0$, $d_2$ the plane of equation $\left( x + \alpha^2 y + \alpha z \right) = 0$
we denote $C_i = d_i \cap H_{\infty}$ for $i = 1 \nocomma, 2$. Let $A_0$ be
the line of equation $x + y = 0$ in $H_{\infty}$ , $A_1$ the line of equation
$y + z = 0$ in $H_{\infty}$ and $A_2$ the line of equation $x + z = 0$ in
$H_{\infty}$.

Let us consider $D$ the divisor associated to the hyperplane section $\bar{X}
\cap H_{\infty}$, so
\[ D = 4 C_1 + 4 C_2 + 3 A_0 + 3 A_1 + 3 A_2 \]
We now denote $\mathfrak{X}_0$ the divisor associated to the hyperplane
section of $X_0$ which is a sub-divisor of $D$ of degree at least 2. We will
denote $\mathfrak{X}_1$ the divisor obtained from $\mathfrak{X}_0$ by applying
the permutation $\left( x, y, z \right)$, $\mathfrak{X}_2$ the divisor
obtained from $\mathfrak{X}_0$ by applying the permutation $\left( x, z, y
\right)$, $\mathfrak{X}_3$ the divisor obtained from $\mathfrak{X}_0$ by
applying the transposition $\left( x, y \right)$, $\mathfrak{X}_4$ the divisor
obtained from $\mathfrak{X}_0$ by applying the transposition $\left( x, z
\right)$ and $\mathfrak{X}_5$ the divisor obtained from $\mathfrak{X}_0$ by
applying the transposition $\left( y, z \right)$. As $\phi \left( x, y, z
\right)$ is symmetrical in $x, y$ and $z$ we know that $\mathfrak{X}_i$ is a
subdivisor of $D$ for $i = 1, \ldots, 5$. The cases where $\mathfrak{X}_0
\geqslant 2 A_0$ or $\mathfrak{X}_0 = A_0$ are already treated in [14] so we
have to study the cases below.

\subsection{Case where $\mathfrak{X}_0$ is of degree 2.}

\begin{enumerateroman}
  \item If $\mathfrak{X}_0 = A_0 + A_1$ therefore from [14] 5.7 we have a
  contradiction with the fact that $\mathfrak{X}_0$ is at most of degree 2.
  
  \item If $\mathfrak{X}_0 = A_0 + C_i$, then $\mathfrak{X}_1 = A_1 + C_i$,
  $\mathfrak{X}_2 = A_2 + C_i$, $\mathfrak{X}_3 = A_0 + C_j$, $\mathfrak{X}_4
  = A_1 + C_j$, $\mathfrak{X}_5 = A_2 + C_j$ with $j \neq i$. As seen in [14]
  the group $< \rho > = \tmop{Gal} \left( \mathbbm{F}_{q^3} /\mathbbm{F}_q
  \right)$ acts on $X_0$ and as $X_0$ is not defined over $\mathbbm{F}_q$
  there exist sub-varieties $X_6$, $X_7$ and $X_8$ which have, respectively
  the associated divisor $\mathfrak{X}_6$, $\mathfrak{X}_7$ and
  $\mathfrak{X}_8$. We have $\mathfrak{X}_6 = A_0 + C_i$, $\mathfrak{X}_7 =
  A_1 + C_i$ and $\mathfrak{X}_8 = A_2 + C_i$. Finally we have $\sum
  \mathfrak{X}_i \geqslant D$ which is a contradiction.
\end{enumerateroman}

\subsection{Case where $\mathfrak{X}_0$ is of degree 3.}

\begin{enumerateroman}
  \item The case where $\mathfrak{X}_0 = A_0 + A_1 + A_2$ has already been
  treated in [14], this is the case where $A + P_1$ divides $\phi$.
  
  \item If $\mathfrak{X}_0$ contains 2 of the $A_i$ from [14] 5.7 it contains
  the 3 and it is the same case than previously.
  
  \item If $\mathfrak{X}_0 = A_0 + 2 C_i$, \ then $\mathfrak{X}_1 = A_1 + 2
  C_i$ and $\mathfrak{X}_2 = A_2 + 2 C_i$, in this case $\mathfrak{X}_0
  +\mathfrak{X}_1 +\mathfrak{X}_2 \geqslant D$ which is a contradiction.
  
  \item If $\mathfrak{X}_0 = A_0 + C_1 + C_2$, then $\mathfrak{X}_1 = A_1 +
  C_1 + C_2$ and $\mathfrak{X}_2 = A_2 + C_1 + C_2$, $\mathfrak{X}_3 = A_0 +
  C_1 + C_2$, $\mathfrak{X}_4 = A_1 + C_1 + C_2$ and $\mathfrak{X}_5 = A_2 +
  C_1 + C_2$. Then $\sum \mathfrak{X}_i \geqslant D$ which is a contradiction.
\end{enumerateroman}

\subsection{Case where $\mathfrak{X}_0$ is of degree 4.}

\begin{enumerateroman}
  \item If $\mathfrak{X}_0 = A_0 + A_1 + A_2 + C_i$, then $\mathfrak{X}_1 =
  A_0 + A_1 + A_2 + C_i$, $\mathfrak{X}_2 = A_0 + A_1 + A_2 + C_i$,
  $\mathfrak{X}_3 = A_0 + A_1 + A_2 + C_j$, $\mathfrak{X}_4 = A_0 + A_1 + A_2
  + C_j$ and $\mathfrak{X}_5 = A_0 + A_1 + A_2 + C_j$. Then $\sum
  \mathfrak{X}_i \geqslant D$ which is a contradiction.
  
  \item If $\mathfrak{X}_0$ contains 2 of the $A_i$ from [14] 5.7 it contains
  the 3 and we are in the same case than in i).
  
  \item If $\mathfrak{X}_0 = A_0 + 3 C_i$, then $\mathfrak{X}_1 = A_1 + 3 C_i$
  and $\mathfrak{X}_2 = A_2 + 3 C_i$. Then $\sum \mathfrak{X}_i \geqslant D$
  which is a contradiction.
  
  \item If $\mathfrak{X}_0 = A_0 + 2 C_i + C_j$ then $\mathfrak{X}_1 = A_1 + 2
  C_i + C_j$ and $X_2 = A_2 + 2 C_i + C_j$, with $j \neq i$. Then $\sum
  \mathfrak{X}_i \geqslant D$ which is a contradiction.
\end{enumerateroman}

\subsection{Case where $\mathfrak{X}_0$ is of degree 5.}

\begin{enumerateroman}
  \item If $X_0 = A_0 + 2 \left( C_1 + C_2 \right)$, then $\mathfrak{X}_1 =
  A_1 + 2 \left( C_1 + C_2 \right)$ and $\mathfrak{X}_2 = A_2 + 2 \left( C_1 +
  C_2 \right)$. Then $\sum \mathfrak{X}_i \geqslant D$ which is a
  contradiction.
  
  \item If $\mathfrak{X}_0 = A_0 + 3 C_i + C_j$, $j \neq i$, $\mathfrak{X}_1 =
  A_1 + 3 C_i + C_j$ and $\mathfrak{X}_2 = A_2 + 3 C_i + C_j$. Then $\sum
  \mathfrak{X}_i \geqslant D$ which is a contradiction.
  
  \item If $\mathfrak{X}_0 = A_0 + 4 C_i$, then $\mathfrak{X}_1 = A_1 + 4 C_i$
  \ then $\mathfrak{X}_0 +\mathfrak{X}_1 \geqslant D$ which is a
  contradiction.
  
  \item If $\mathfrak{X}_0$ contains 2 of the $A_i$ from [14] 5.7 it contains
  the 3 and we will treat those cases in the following points.
  
  \item If $\mathfrak{X}_0 = A_0 + A_1 + A_2 + 2 C_i$ then $\mathfrak{X}_1 =
  A_0 + A_1 + A_2 + 2 C_i$ and $\mathfrak{X}_2 = A_0 + A_1 + A_2 + 2 C_i$.
  Then $\sum \mathfrak{X}_i \geqslant D$ which is a contradiction.
  
  \item The only case left is when $X_0 = A_0 + A_1 + A_2 + C_1 + C_2$. As
  seen in [14] the group $< \rho > = \tmop{Gal} \left( \mathbbm{F}_{q^3}
  /\mathbbm{F}_q \right)$ acts on $X_0$ and as $X_0$ is not defined over
  $\mathbbm{F}_q$ there exist sub-varieties $X_6$, $X_7$ and $X_8$ which have,
  respectively the associated divisor $\mathfrak{X}_6$, $\mathfrak{X}_7$ and
  $\mathfrak{X}_8$. We have $\mathfrak{X}_6 = A_0 + A_1 + A_2 + C_1 + C_2$,
  $\mathfrak{X}_7 = A_0 + A_1 + A_2 + C_1 + C_2$ and $\mathfrak{X}_8 = A_0 +
  A_1 + A_2 + C_1 + C_2$. It remains the sub-divisor $\mathfrak{X}_9 = C_1 +
  C_2$. Therefore $\sum \mathfrak{X}_i = D$ and the form of $\phi$ is :
  \[ \phi = \left( \phi_5 + R \right) \left( A \phi_5 + Q \right) \left( A
     \phi_5 + \rho \left( Q \right) \right) \left( A \phi_5 + \rho^2 \left( Q
     \right) \right) \]
  with $R$ a polynomial of degree 1 such as $\phi_5 + R$ is not irreducible,
  $Q$ a polynomial of degree 4 and $\rho$ the generator of $\tmop{Gal} \left(
  \mathbbm{F}_{q^3} /\mathbbm{F}_q \right)$.
\end{enumerateroman}
It is useless to consider the cases where $X_0$ is of degree more than 5 as we
obtain 2 other divisors of the same degree from $X_0$ and $D$ is of degree 17.
Therefore it is sufficient to prove the theorem 7.

\section{Proof of theorem 8}

We have the two following cases to study:

\subsection{Case where $A + P_1$ divides $\phi$.}

If $P_1$ divides $\phi$ then $\left( A + P_1 \right) \left( A + \rho \left(
P_1 \right) \right) \left( A + \rho^2 \left( P_1 \right) \right)$ divides
$\phi$ too (see [14]. By calculus (see Appendix 1) we can state that:
\begin{itemizedot}
  \item $P_1 = c_1 \phi_5 + c_1^3$.
  
  \item The trace of $c_1$ in $\mathbbm{F}_{q^3}$ is 0.
  
  \item $\left( A + P_1 \right) \left( A + \rho \left( P_1 \right) \right)
  \left( A + \rho^2 \left( P_1 \right) \right)$ is the polynomial $\phi$
  associated to $L \left( x \right)^3$ where $L \left( x \right) = x \left( x
  + c_1 \right) \left( x + \rho \left( c_1 \right) \right) \left( x + \rho^2
  \left( c_1 \right) \right)$.
  
  \item $\tmop{We} \tmop{have} f = L \left( x \right)^3 \left( L \left( x
  \right)^2 + a \right) + a_{16} x^{16} + a_8 x^8 + a_4 x^4 + a_2 x^2 + a_1 x
  + a_0$ where $a, a_0, a_1, a_2, a_4, a_8, a_{16} \in \mathbbm{F}_q$.
\end{itemizedot}

By proposition 9 $f$ is equivalent to $L \left( x \right)^5 + aL \left( x
\right)^3$. As $\tmop{tr} \left( c_1 \right) = 0$, $L \left( x \right)$ is a
$q$-affine permutation hence $f$ is CCZ-equivalent to $x^5 + ax^3$.

By theorem 3 $f$ cannot be APN over infinitely many extensions of
$\mathbbm{F}_q$ if $a \neq 0$. Hence $a = 0$ and $f$ is CCZ-equivalent to
$x^5$, which is a gold function.

\subsection{Case where $P_2$ divides $\phi$.}

If $P_2$ divides $\phi$ then, by calculus (see Appendix 2), we obtain that $f
= \left( x^{20} + ax^{10} + bx^5 \right) + a_{16} x^{16} + a_8 x^8 + a_4 x^4 +
a_2 x^2 + a_1 x + a_0$, where $a, b, a_0, a_1, a_2, a_4, a_8, a_{16} \in
\mathbbm{F}_q$. By proposition 9 $f$ is equivalent to $\left( x^{5} + ax^2 +
bx \right)^4$. Therefore $f$ can be written $f \left( x \right) = L \left( x^5
\right)$ with $L \left( x \right) = x^{4} + ax^2 + bx$ which is a
permutation. Hence, $f$ is CCZ-equivalent to $x^5$.

In conclusion, we proved that if $f \left( x \right)$ is a polynomial of
$\mathbbm{F}_q$ of degree 20 which is APN over infinitely many extensions of
$\mathbbm{F}_q$, then $f \left( x \right)$ is CCZ-equivalent to $x^5$.

\section{Appendix}

In this part we give the details of the calculus we made in order to state the
theorem 8 We just use the fact that $P_1$ or $P_2$ divides $\phi$ and it gives
us conditions on the coefficients of $P_1$ or $P_2$ and $\phi$. As $\phi$ is a
symmetrical polynomial in $x, y, z$ we can write it using symmetrical
functions $s_1 = x + y + z$, $s_2 = xy + xz + yz$ and $s_3 = xyz$. We recall
that $\phi_i$ is the polynomial $\phi$ associated to $x^i$ and therefore $\phi
\left( x, y, z \right)$ the polynomial associated to $f \left( x \right) =
\sum_{i = 0}^d a_i x^i$ can be written $\phi = \sum a_i \phi_i$. Denoting $p_i
= x^i + y^i + z^i$, we have $p_i = s_1 p_{i - 1} + s_2 p_{i - 2} + s_3 p_{i -
3} \nocomma$. We remark that $\phi_i = \frac{p_i + s_1^i}{A}$ and that $A =
s_1 s_2 + s_3$.

The calculus were made on the software Sage and you can find the sheet at the
following adress: \href{}{http://sagenb.org/home/pub/5035}.

\subsection{Case where $A + P_1$ divides $\phi$.}

We will write $P$ for $P_1$ in this section in order to make the calculus more
readable.

If $A + P$ divides $\phi$ then $\left( A + P \right) \left( A + \rho \left( P
\right) \right) \left( A + \rho^2 \left( P \right) \right)$ is of degree 9 and
divides $\phi$ too (see [14]). We write
\[ \left( A + P \right) \left( A + \rho \left( P \right) \right) \left( A +
   \rho^2 \left( P \right) \right) = \sum_{i = 0}^9 P_i, \]
where $P_i$ is the term of degree $i$ of $\left( A + P_{} \right) \left( A +
\rho \left( P \right) \right) \left( A + \rho^2 \left( P \right) \right)$.

As $\left( A + P \right) \left( A + \rho \left( P \right) \right) \left( A +
\rho^2 \left( P \right) \right)$ divides $\phi$ there exists a polynomial $Q$
of degree 8 such as
\[ \phi = \left( A + P \right) \left( A + \rho \left( P \right) \right)
   \left( A + \rho^2 \left( P \right) \right) Q \]
and we write
\[ Q = \sum_{i = 0}^8 Q_i, \]
where $Q_i$ is the term of degree $i$ of $Q$.

\subsubsection{Degree 17}

We put $a_{20} = 1$ and we have :
\[ \phi_{20} = P_9 Q_8 . \]
As $P_9 = A^3$ we have $Q_8 = \phi_5^4$.

\subsubsection{Degree 16.}

We have
\[ a_{19} \phi_{19} = P_9 Q_7 + P_8 Q_8 . \]
As $P_8 = A^2 (s_1^2 \tmop{tr} (c_1) + s_2 \tmop{tr} (c_4))$, where $\tmop{tr}
\left( c_i \right)$ is the trace of $c_1$, it gives us
\[ a_{19} \phi_{19} = A^3 Q_7 + A^2 \phi_5^4  (s_1^2 \tmop{tr} (c_1) + s_2
   \tmop{tr} (c_4)) . \]
As $\phi_{19}$ is not divisible by $A$ (by lemma 13) so $a_{19} = 0$ and
\[ AQ_7 = \phi_5^4 (s_1^2 \tmop{tr} (c_1) + s_2 \tmop{tr} (c_4)) . \]
We know that $A$ is prime with $s_1^2 \tmop{tr} (c_1) + s_2 \tmop{tr} (c_4)$
because $(x + y)$ does not divide this polynomial, and $A$ does not divide
either $\phi_5^4$ which implies $Q_7 = P_8 = 0$ and $\tmop{tr} (c_1) =
\tmop{tr} (c_4) = a_{19} = 0$.

\subsubsection{Degree 15.}

We have
\[ a_{18} \phi_{18} = a_{18} (A \phi_9^2) = P_9 Q_6 + P_8 Q_7 + P_7 Q_8 . \]
Knowing that $P_8 = Q_7 = 0$ we obtain
\[ a_{18} (A \phi_9^2) = P_9 Q_6 + P_7 Q_8 = A^3 Q_6 + \phi_5^4 P_7 . \]
We also know that

\begin{flushleft}
  \begin{center}
    \begin{flushright}
      \begin{flushleft}
        \[ \phi_5^4 = \left( s_1^2 + s_2 \right)^4 = s_1^8 + s_2^4  \]
      \end{flushleft}
    \end{flushright}
  \end{center}
\end{flushleft}

and

\ \ \ \ \ \ \ \ \ \ \ \ \ \ $P_7 = A \left( s_1^4 q_1 \left( c_1 \right) +
s_2^2 q_1 \left( c_4 \right) + s_1^2 s_2 q_5 \left( c_1, c_4 \right) \right) +
A^2 s_1 \tmop{tr} \left( b_1 \right)$

denoting

$q_1 \left( c_i \right) = c_i \rho \left( c_i \right) + c_i \rho^2 \left( c_i
\right) + \rho \left( c_i \right) \rho^2 \left( c_i \right)$ and

$q_5 (c_1, c_4) = c_1 \left( \rho (c_4) + \rho^2 (c_4) \right) + c_4 \left(
\rho (c_1) + \rho^2 (c_1) \right) + \rho (c_1) \rho^2 (c_4) + \rho (c_4)
\rho^2 \left( c_1 \right)$.

So
\[ a_{18} \phi_9^2 = A^2 Q_6 + \phi_5^4 \left( s_1^4 q_1 \left( c_1 \right) +
   s_2^2 q_1 \left( c_4 \right) + s_1^2 s_2 q_5 \left( c_1, c_4 \right)
   \right. + As_1 \left. \tmop{tr} \left( b_1 \right) \right), \]
hence $A$ divide $a_{18} \phi_9^2 + \phi_5^4 \left( s_1^4 q_1 \left( c_1
\right) + s_2^2 q_1 \left( c_4 \right) + s_1^2 s_2 q_5 \left( c_1, c_4 \right)
\right)$. As $A = s_1 s_2 + s_3$ the polynomial $a_{18} \phi_9^2 + \phi_5^4
\left( s_1^4 q_1 \left( c_1 \right) + s_2^2 q_1 \left( c_4 \right) + s_1^2 s_2
q_5 \left( c_1, c_4 \right) \right)$ cannot contain monomial in $s_1^{12}$ or
$s_2^6$, therefore $a_{18} = q_1 \left( c_1 \right) = q_1 \left( c_4 \right)$.

Then $A$ divides $a_{18} \left( \phi_9^2 + \phi_5^6 \right) + \phi_5^4 s_1^2
s_2 q_5 \left( c_1, c_4 \right)$. As $\phi_9^2 + \phi_5^6 = A^4$ and $A$ does
not divide $\phi_5^4$ we have $q_5 \left( c_1, c_4 \right) = 0$. Replacing in
the first equation we have
\[ a_{18} A^4 = A^2 Q_6 + A \phi_5^4 s_1 \left( \tmop{tr} \left( b_1 \right)
   \right) . \]
So
\[ a_{18} A^3 + AQ_6 = \phi_5^4 s_1 \left( \tmop{tr} \left( b_1 \right)
   \right), \]
as $A$ does not divide $\phi_5^4 s_1$, $\tmop{tr} \left( b_1 \right) = 0$ and
$Q_6 = a_{18} A^2$.

\subsubsection{Degree 14.}

\tmtextbf{We first prove that $c_1 = c_4$.}

We have
\[ a_{17} \phi_{17} = P_9 Q_5 + \ldots + P_6 Q_8 = P_9 Q_5 + P_6 Q_8 . \]
We know that
\[ P_6 = A^2 N (d) + A \left( s_1^3 q_5 (c_1, b_1) + s_1 s_2 q_5 (c_1, b_1)
   \right) + s_1^6 N (c_1) + s_1^4 s_2 q_4 (c_1, c_4) + s_1^2 s_2^2 q_4 (c_4,
   c_1) + s_2^3 N (c_4) \]
where
\[ 
 N (a) = a \rho (a) \rho^2 (a) \text{which is the norm of} a \text{in} \mathbbm{F}_q .\\
   q_4 (a, b) = a \rho (a) \rho^2 (b) + a \rho (b) \rho^2 (a) + b \rho (a)\rho^2 (a). \\   
   q_5 (a, b) = a (\rho (b) + \rho^2 (b)) + b (\rho (a) + \rho^2 (a)) + \rho(a) \rho^2 (b) + \rho (b) \rho^2 (a) .
 \]
for all $a, b$ in $\mathbbm{F}_{q^3}$.

We can write
\[ P_6 = A^2 \tmop{tr} (d) + A \left( s_1^3 q_5 (c_1, b_1) + s_1 s_2 q_5
   (c_1, b_1) \right) + P_6^{\ast} \rho (P_6^{\ast}) \rho^2 (P_6^{\ast}), \]
where $P_6^{\ast} = c_1 s_1^2 + c_4 s_2$. So we can deduce that
\[ a_{17} \phi_{17} = A^3 Q_5 + \phi_5^4 \left( A^2 \tmop{tr} (d) + A \left(
   s_1^3 q_5 (c_1, b_1) + s_1 s_2 q_5 (c_1, b_1) \right) + P_6^{\ast} \rho
   (P_6^{\ast}) \rho^2 (P_6^{\ast}) \right) . \]
We now have $A$ divides $a_{17} \phi_{17} + \phi_5^4 P_6^{\ast} \rho
(P_6^{\ast}) \rho^2 (P_6^{\ast})$. In addition, denoting $s = x + y$,
\[ (x + z)^2 \phi_5 = (x + z)^4 + s (x^2 y + x^2 z + yz^2 + z^3) = (x + z)^4 +
   sR_1 \]
and
\[ (x + z)^2 \phi_{17} = (x + z)^{16} + sR_2, \]
where $R_1$ is a polynomial of degree 3 and $R_2$ is a polynomial of degree
15. As $(x + z)^8 A = (x + z)^9 s (x + z + s)$ divides $a_{17} (x + z)^8
\phi_{17} + P_6^{\ast} \rho (P_6^{\ast}) \rho^2 (P_6^{\ast}) (x + z)^8
\phi_5^4 $ which is equal to
\begin{equation}
  a_{17} (x + z)^6 \left( x^{16} + z^{16} + sR_1 \right) + P_6^{\ast} \rho
  (P_6^{\ast}) \rho^2 (P_6^{\ast}) \left( x^4 + z^4 + sR_2 \right) .
\end{equation}
Therefore we have $P_6^{\ast} = c_1 (s^2 + z^2) + c_4 (x^2 + s (x + z)) = c_1
z^2 + c_4 x^2 + sR_3 = P_6^{\ast \ast} + sR_3$.

As $s$ divides (1) the constant term in $s$ vanishes :
\[ (x + z)^{16} \left( a_{17} (x + z)^6 + P_6^{\ast \ast} \rho (P_6^{\ast
   \ast}) \rho^2 (P_6^{\ast \ast}) \right) = 0, \]
then
\[ a_{17} (x + z)^6 + P_6^{\ast \ast} \rho (P_6^{\ast \ast}) \rho^2 (P_6^{\ast
   \ast}) = 0, \]
hence
\[ a_{17} (x + z)^6 + (c_4 x^2 + c_1 z^2) (\rho (c_4) x^2 + \rho (c_1) z^2)
   (\rho^2 (c_4) x^2 + \rho^2 (c_1) z^2) = 0, \]
so
\[ a_{17} (x + z)^3 + ( \sqrt{c_4} x + \sqrt{c_1} z) (\rho ( \sqrt{c_4}) x +
   \rho ( \sqrt{c_1}) z) (\rho^2 ( \sqrt{c_4}) x + \rho^2 ( \sqrt{c_1}) z) =
   0. \]
The polynomial $x + z$ divides $( \sqrt{c_4} x + \sqrt{c_1} z) (\rho (
\sqrt{c_4}) x + \rho ( \sqrt{c_1}) z) (\rho^2 ( \sqrt{c_4}) x + \rho^2 (
\sqrt{c_1}) z)$ so it divides one component and then $c_1 = c_4$.

\tmtextbf{We now calculate $P_6$ and $Q_5$.}

As $c_1 = c_4$ we have $P_6 = A^2 \tmop{tr} \left( d \right) + A \phi_5 s_1
q_5 \left( c_1, b_1 \right) + \phi_5^3 N \left( c_1 \right)$, so from $a_{17}
\phi_{17} = P_9 Q_5 + P_6 Q_8$ we can deduce that $A$ divides $a_{17}
\phi_{17} + q_3 \left( c_1 \right) \phi_5^7$. Hence the coefficient of the
monomials $s_1^{14}$ in $a_{17} \phi_{17} + N \left( c_1 \right) \phi_5^7$,
which is $a_{17} + N \left( c_1 \right)$, must be equal to 0, so $a_{17} = N
\left( c_1 \right)$.

Remarking that $\phi_{17} + \phi_5^7 = A^2 \phi_5 \phi_9$ we have $A$ divides
$\phi_5 s_1 q_5 \left( c_1, b_1 \right)$. As $\phi_5 s_1$ is not divisible by
$A$ we have $q_5 \left( c_1, b_1 \right)$=0. So now we have
\[ A^3 Q_5 = A^2 \phi_5^4 \tmop{tr} \left( d \right) + a_{17} A^2 \phi_5
   \phi_9, \]
which gives
\begin{eqnarray*}
  & AQ_5 = \phi_5^4 \tmop{tr} \left( d \right) + a_{17} \phi_5 \phi_9 . & 
\end{eqnarray*}
Using the same argument as precedent we have $\tmop{tr} \left( d \right) = N
\left( c_1 \right)$ and then $Q_5 = a_{17} \frac{\phi_5 \phi_9 +
\phi_5^4}{A^2} = a_{17} A^2 \phi_5$ and $P_6 = a_{17} \left( A^2 + \phi_5^3
\right)$.

\subsubsection{Degree 13}

We use
\[ 0 = a_{16} \phi_{16} = P_9 Q_4 + P_8 Q_5 + P_7 Q_6 + P_6 Q_7 + P_5 Q_8 \]
\[ = A^3 Q_4 + a^2_{18} A^3 \phi_5^2 + \phi_5^4 P_5, \]
with
\[ P_5 = q_4 (c_1, b_1) \left( s_1^5 + s_2^2 s_1 \right) + A \left( q_1 (b_1)
   s_1^2 + q_5 (c_1, d) \left( s_1^2 + s_2 \right) \right) . \]
As $A^3$ does not divide $P_5 \phi_5^4$, so $P_5 = 0$ and $q_4 (c_1, b_1) =
q_1 (b_1) = q_5 (c_1, d) = 0$. We deduce
\[ Q_4 = a^2_{18} \phi_5^2 . \]

\subsubsection{Degree 12}

We have
\[ a_{15} \phi_{15} = P_9 Q_3 + P_8 Q_4 + P_7 Q_5 + P_6 Q_6 + P_5 Q_7 + P_4
   Q_8 \]
\[ = A^3 Q_3 + a_{18} a_{17} A^2 \phi_5^3 + a_{18} a_{17} A^2 \left( A^2 +
   \phi_5^3 \right) + P_4 \phi_5^4, \]
with
\[ P_4 = q_4 (b_1, c_1) (s_1^4 + s_1^2 s_2) + q_4 (c_1, d) (s_1^4 + s_2^2) +
   q_5 (b_1, d) As_1 = H_4 + AG_4, \]
where $H_4 = q_4 (b_1, c_1) (s_1^4 + s_1^2 s_2) + q_4 (c_1, d) (s_1^4 +
s_2^2)$ and $G_4 = q_5 (b_1, d) s_1$. So $A | H_4 \phi_5^4 + a_{15}
\phi_{15}$. As
\[ H_4 \phi_5^4 + a_{15} \phi_{15} = s_1^{12} \left( a_{15} + q_4 (b_1, c_1) +
   q_4 (c_1, d) \right) + s_1^{10} s_2 q_4 (b_1, c_1) + a_{15} s_1^9 s_3 +
   s_1^8 s_2^2  \left( a_{15} + q_4 (c 1, d) \right) + s_1^4 s_2^4  \left( q_4
   (b_1, c_1) + q_4 (c_1, d) \right) + s_1^2 s_2^5 q_4 (b 1, c 1) + a_{15}
   \left( s_1^3 s_3^3 + s_1 s_2^4 s_3 + s_3^4 \right) + s_2^6  \left( a_{15} +
   q_4 (c 1, d) \right), \]
the coefficients of $s_1^{12}$ and $s_2^6$ must be 0 and so
\[ a_{15} + q_4 (b 1, c 1) + q_4 (c 1, d) = 0 \text{ and}\\
   a_{15} + q_4 (c 1, d) = 0 \tmop{so} q_4 (b 1, c 1) = 0. \]
Replacing in the equation we now have
\[ H_4 \phi_5^4 + a_{15} \phi_{15} = a_{15} \left( s_1^9 s_3 + s_1^4 s_2^4 +
   s_1^3 s_3^3 + s_1 s_2^4 s_3 + s_3^4 \right) = a_{15} \left( \phi_{15} +
   \phi_5^6 \right), \]
but $A$ does not divide $\phi_{15} + \phi_5^6$ so $a_{15} = 0$ so $H_4 = 0$.

Hence
\[ 0 = A^3 Q_3 + a_{18} a_{17} A^2 \phi_5^3 + a_{18} a_{17} A^2 \left( A^2 +
   \phi_5^3 \right) + AG_4 \phi_5^4 . \]
So $A$ divides $G_4$, but the degree of $G_4$ is less than or equal to 1 so
$G_4 = 0$ it implies $q_5 (b_1, d) = 0$ so $P_4 = 0$.

We conclude
\[ Q_3 = a_{18} a_{17} A. \]

\subsubsection{Degree 11.}

We have
\[ a_{14} \phi_{14} = P_9 Q_2 + P_8 Q_3 + P_7 Q_4 + P_6 Q_5 + P_5 Q_6 + P_4
   Q_7 + P_3 Q_8, \]
so
\[ a_{14} A (\phi_5^4 + s_1^2 s_3^2) = A^3 Q_2 + a_{18}^3 A \phi_5^4 +
   a_{17}^2 A \phi_5 \left( A^2 + \phi_5^3 \right) + P_3 \phi_5^4 . \left(
   \ast \right) \]
So $A$ divides $P_3$. But $P_3 = N (b_1) s_1^3 + q_6 (c_1, b_1, d)) s_1 \phi_5
+ q_1 (d) A$ so $N (b_1) = q_6 (c_1, b_1, d) = 0$ with
\[ q_6 (c_1, b_1, d) = b_1 \rho (c_1) \rho^2 (d) + b_1 \rho (d) \rho^2 (c_1) +
   c_1 \rho (b_1) \rho^2 (d) + c_1 \rho (d) \rho^2 (b_1) + d \rho (c_1) \rho^2
   (b_1) + d \rho (c_1) \rho^2 (b_1) . \]

As $N \left( b_1 \right) = 0, b_1 = 0$.

When we replace in the equation $\left( \ast \right)$ we have
\[ A^3 (Q_2 + a_{17}^2 \phi_5) = A \left( \phi_5^4 \left( a_{14} + a_{18}^3 +
   a_{17}^2 + q_1 (d) \right) + a_{14} s_1^2 s_3^2 \right), \]
so $A$ divides $\phi_5^4 \left( a_{14} + a_{18}^3 + a_{17}^2 + q_1 (d) \right)
+ a_{14} s_1^2 s_3^2 = (s_1^8 + s_2^4) \left( a_{14} + a_{18}^3 + a_{17}^2 +
q_1 (d) \right) + a_{14} s_1^2 s_3^2$, then $a_{14} + a_{18}^3 + a_{17}^2 +
q_1 (d) = 0$, with the same argument as before on the coefficients of the
monomials $s_1^8$ and $s_2^4$, therefore $a_{14} = 0$ because $A$ does not
divide $s_1^2 s_3^2$.

We obtain
\[ Q_2 = a^2_{17} \phi_5, \]
and
\[ P_3 = (a_{17}^2 + a_{18}^3) A. \]

\subsubsection{Degree 10.}

We have
\[ a_{13} \phi_{13} = P_9 Q_1 + P_8 Q_2 + P_7 Q_3 + P_6 Q_4 + P_5 Q_5 + P_4
   Q_6 + P_3 Q_7 + P_2 Q_8 \]
\[ = A^3 Q_1 + a_{17} a^2_{18} A^2 \phi_5^2 + a_{17} a^2_{18} \phi_5^2 \left(
   A^2 + \phi_5^3 \right) + \phi^4_5 \left( s_1^2 q_4 \left( d, c 1 \right) +
   s_2 q_4 \left( d, c 1 \right) \right), \]
so $A$ divides $a_{13} \phi_{13} + \phi_5^5 \left( a_{17} a^2_{18} + q_4
\left( d, c 1 \right) \right) = a_{13} \left( s_1^4 s_3^2 + s_1^3 s_2^2 s_3 +
s_1^2 s_2 s_3^2 + s_1 s_3^3 \right) + \phi_5^5 \left( a_{13} + a_{17} a_{18}^2
+ q_4 \left( d, c 1 \right) \right)$. with the same argument as before on the
coefficients of the monomials $s_1^8$ and $s_2^4$ we have
\[a_{13} + a_{17} a_{18}^2 + q_4 \left( d, c 1 \right) = 0. \]
in addition, $A$ does not divide $s_1^4 s_3^2 + s_1^3 s_2^2 s_3 + s_1^2 s_2
s_3^2 + s_1 s_3^3$ so $a_{13} = 0$ and $q_4 (d, c 1) = a_{17} a^2_{18}$.

Now we have
\[ AQ_1 = 0. \]
So $Q_1 = 0$ and $P_2 = a_{17} a_{18}^2 \phi_5$.

\subsubsection{Degree 9.}

We have
\[ a_{12} \phi_{12} = P_9 Q_0 + P_8 Q_1 + P_7 Q_2 + P_6 Q_3 + P_5 Q_4 + P_4
   Q_5 + P_3 Q_6 + P_2 Q_7 + P_1 Q_8, \]
but $\phi_{12} = A^3$ and as $b_1 = 0$ we have $P_1 = 0$. So
\[ a_{12} A^3 = A^3 Q_0 + a_{17}^2 a_{18}^{} A \phi_5^3 + a_{17}^2 a_{18}^{}
   A \left( A^2 + \phi_5^3 \right) + a_{18} (a^2_{17} + a_{18}^3) A^3, \]
so $Q_0 = a_{12} + a_{18}^4$.

\subsubsection{Degree 8.}

We have
\[ a_{11} \phi_{11} = P_8 Q_0 + P_7 Q_1 + P_6 Q_2 + P_5 Q_3 + P_4 Q_4 + P_3
   Q_5 + P_2 Q_6 + P_1 Q_7 + P_0 Q_8, \]
which gives
\[ a_{11} \phi_{11} = (P_0 + a_{17}^3) \phi_5^4 . \]
But $\phi_5$ does not divide $\phi_{11}$ so $a_{11} = 0$ et $P_0 = a_{17}^3$.

\subsubsection{Conclusion.}

We now have the following systems:
\[ \left\{ \begin{array}{l}
     \tmop{tr} \left( c_1 \right) = 0\\
     N (c_1) + \tmop{tr} (d) = 0\\
     q_5 (c_1, d) = 0\\
     q_4 (c_1, d) = 0\\
     q_1 (d) = q_1^3 (c_1) + N (c_1)^2\\
     q_4 (d, c_1) = N (c_1) q_1^2 (c_1)\\
     N (d) = N (c_1)^3
   \end{array} \right. \]
and
\[ a_{18} = q_1 (c_1), a_{17} = N (c_1) = \tmop{tr} (d) . \]
Solving the system formed by the linear equations in $d, \rho (d), \rho^2
(d)$, we obtain $d = c_1^3$. We also have $b_1 = 0$ as $b_1 \rho \left( b_1
\right) \rho^2 \left( b_1 \right) = 0$. Therefore
\[ P = c_1 \phi_5 + c_1^3, \]
and
\[ Q = \phi_5^4 + q_1 (c_1) A^2 + N (c_1) A \phi_5 + q_1 (c_1)^2 \phi_5^2 +
   q_1 (c_1) N (c_1) A + q_3 (c_1)^2 \phi_5 + a_{12} + q_1 (c_1)^4, \]
therefore
\[ f \left( x \right) = x^{20} + a_{18} x^{18} + a_{17} x^{17} + a_{16}
   x^{16} + a_{12} x^{12} + a_{18} a_{12} x^{10} + a_{17} a_{12} x^9 + a_8 x^8
   + \left( a_{18}^7 + a_{18}^4 a_{17}^2 + a_{18}^3 a_{12} + a_{18} a_{17}^4 +
   a_{17}^2 a_{12} \right) x^6 + \left( a_{18}^6 a_{17} + a_{18}^2 a_{17}
   a_{12} + a_{17}^5 \right) x^5 + a_4 x^4 + \left( a_{18}^4 a_{17}^3 +
   a_{17}^3 a_{12} \right) x^3 + a_2 x^2 + a_1 x + a_0 . \]
Putting $L \left( x \right) = x \left( x + c_1 \right) \left( x + \rho \left(
c_1 \right) \right) \left( x + \rho^2 \left( c_1 \right) \right)$ we have that
$\left( A + P \right) \left( A + \rho \left( P \right) \right) \left( A +
\rho^2 \left( P \right) \right)$ is the polynomial $\phi$ associated to $L
\left( x \right)^3$ wich leads us to study the divisibility of $f$ by $L
\left( x \right)^3$. We have in our case $f = L \left( x \right)^3 \left( L
\left( x \right)^2 + a_{12} \right) + a_{16} x^{16} + a_8 x^8 + a_4 x^4 + a_2
x^2 + a_1 x + a_0$.

\subsection{Case where $P_2$ divides $\phi$.}

We will write $P$ for $P_2$ in this section in order to make the calculus more
readable.

From theorem 7 we have
\[ \phi = \left( \phi_5 + R \right) \left( A \phi_5 + Q \right) \left( A
   \phi_5 + \rho \left( Q \right) \right) \left( A \phi_5 + \rho^2 \left( Q
   \right) \right), \]
where $R$ is a symmetrical polynomial of $\mathbbm{F}_q$ of degree 1 and $Q$
is a symmetrical polynomial of $\mathbbm{F}_{q^3}$ of degree 4. We will denote
$R = as_1 + b$ and $\left( A \phi_5 + Q \right) \left( A \phi_5 + \rho \left(
Q \right) \right) \left( A \phi_5 + \rho^2 \left( Q \right) \right) = \sum_{i
= 0}^{15} Q_i$. We will identify degree by degree the expression of $\phi$.

\subsubsection{Degree 17.}

We have
\[ \phi_{20} = A^3 \phi_5^4 = \phi_5 Q_{15}, \]
so $Q_{15} = A^3 \phi_5^3$.

\subsubsection{Degree 16.}

We have
\[ a_{19} \phi_{19} = \phi_5 Q_{14} + as_1 Q_{15} = \phi_5 Q_{14} + as_1 A^3
   \phi_5^3, \]
which implies $\phi_5$ divides $\phi_{19}$ but this is not the case hence
$a_{19} = 0$ and $Q_{14} = as_1 A^3 \phi_5^2$.

\subsubsection{Degree 15.}

We have
\[ a_{18} \phi_{18} = \phi_5 Q_{13} + as_1 Q_{14} + bQ_{14} = \phi_5 Q_{13} +
   as_1^2 A^3 \phi_5^2 + bA^3 \phi_5^3, \]
which implies $\phi_5$ divides $\phi_{18}$ but this is not the case hence
$a_{18} = 0$ and $Q_{13} = A^3 \left( a^2 s_1^2 \phi_5 + b \phi_5^2 \right)$.

\subsubsection{Degree 14 and 13}

We have
\begin{equation}
  a_{17} \phi_{17} = \phi_5 Q_{12} + as_1 Q_{13} + bQ_{14},
\end{equation}
and
\begin{equation}
  a_{16} \phi_{16} = 0 = \phi_5 Q_{11} + as_1 Q_{12} + bQ_{13} = \phi_5 Q_{11}
  + as_1 Q_{12} + bA^3 \left( a^2 s_1^2 \phi_5 + b \phi_5^2 \right),
\end{equation}
(2) implies that $Q_{12}$ is divisible by $\phi_5$ or $a = 0$. Lets assume $a
\neq 0$. From (2) we have
\[ a_{17} \frac{\phi_{17}}{\phi_5} = Q_{12} + a^3 s_1^3 A^3 . \]
(we can show easily that $\phi_5$ divides $\phi_{17}$ by calculus). As
{\tmem{$\phi_5$}} divides $Q_{12}$ it divides $a_{17} \frac{\phi_{17}}{\phi_5}
+ a^3 s_1^3 A^3$ too. But
\[ a_{17} \frac{\phi_{17}}{\phi_5} + a^3 s_1^3 A^3 = a_{17} s_3^4 + R_1, \]
so $a_{17} = 0$. As $\phi_5$ does not divide $s_1^3 A^3$ it means $a = 0$ and
$Q_{12} = 0$. We now have, in both case
\[ \phi = \left( \phi_5 + b \right) \left( \sum_{i = 0}^{15} Q_i \right) . \]
We know that $\phi_5 + b$ is irreducible if $b \neq 0$ ([11]), which is in
contradiction with the fact that $f$ is APN over infinitely many extension of
$\mathbbm{F}_q$ and then $b = 0$.

We now have $Q_{15} = A^3 \phi_5^3 \nocomma, Q_{14} = Q_{13} = Q_{12} = Q_{11}
= 0.$

\subsubsection{Degree 12 to 8.}

We have
\[ a_{15} \phi_{15} = \phi_5 Q_{10}, \]
as $\phi_5$ does not divide $\phi_{15}$ we have $a_{15} = 0$ and $Q_{10} = 0$.
The same method can be applied until the degree 8. It gives $a_{14} = a_{13} =
a_{12} = a_{11} = 0$ and $Q_9 = Q_8 = Q_7 = Q_6 = 0$.

\subsubsection{Degree 7.}

We have
\[ a_{10} \phi_{10} = a_{10} A \phi_5^2 = Q_5 \phi_{10}, \]
so $Q_5 = a_{10} A \phi_5$.

\subsubsection{Degree 6.}

The same argument than in section 7.2.5 gives $a_9 = 0$ and $Q_4 = 0$.

\subsubsection{Degree 5.}

We have
\[ a_8 \phi_8 = 0 = Q_3 \phi_5, \]
therefore $Q_3 = 0$.

\subsubsection{Degree 4 and 3.}

The same argument than in section 7.2.5 gives $a_7 = a_6 = 0$ and $Q_2 = Q_1 =
0$.

\subsubsection{Degree 2.}

We have
\[ a_5 \phi_5 = Q_0 \phi_5, \]
therefore $Q_0 = a_5$.

\subsubsection{Conclusion.}

In conclusion we have
\[ \phi = \phi_5 \left( A^3 \phi_5^3 + a_{10} A \phi_5 + a_5 \right) =
   \phi_{20} + a_{10} \phi_{10} + a_5 \phi_5, \]
which gives $f \left( x \right) = x^{20} + a_{16} x^{16} + a_{10} x^{10} + a_8
x^8 + a_5 x^5 + a_4 x^4 + a_2 x^2 + a_1 x + a_0$.

\end{document}